\documentclass[preprint,aps,amsmath,nofootinbib,tightenlines,floatfix]{revtex4}
\usepackage{amssymb}
\usepackage{graphicx,epsfig,amssymb}
\usepackage{multirow}

\begin{document}

\title{Sbottom Signature of the Supersymmetric Golden Region}

\author{Honglei Li$^{1}$\footnote{lihl@mail.sdu.edu.cn},
Will Parker$^{2}$\footnote{wcparker@wisc.edu},
        Zongguo Si$^{1}$\footnote{zgsi@sdu.edu.cn}, \,and\,
        Shufang Su$^{3,4}$\footnote{shufang@physics.arizona.edu}}
\affiliation{$^{1}$ Department of Physics, Shandong University, Jinan, Shandong 250100, China\\
$^{2}$ Department of Physics, University of Wisconsin, Madison, Wisconsin 53706, USA\\
$^{3}$ Department of Physics, University of Arizona, Tucson, Arizona  85721 USA\\
$^{4}$  Department of Physics and Astronomy, University of California, Irvine, California 92697, USA}
\begin{abstract}
The experimental search limit on the Higgs boson mass points to a ``golden region" in
the Minimal Supersymmetric Standard Model parameter space in which
the fine-tuning in the electroweak sector is minimized. One of the sbottoms is relatively light  since 
its mass parameter is related to that of the stop.  
The decay products of the light sbottom typically include
a $W$ boson and a $b$ jet.   We studied the pair production of the light sbottoms at the
Large Hadron Collider  and examined its discovery potential via
collider signature of $4$ jets $+$ 1 lepton $+$ missing $E_T$.  We
analyzed the Standard Model backgrounds for this channel and
developed a set of cuts to identify the signal and suppress the
backgrounds.  We showed that with 100 ${\rm fb}^{-1}$ integrated
luminosity, a significant level of 5 $\sigma$ could be reached for
the light sbottom discovery at the LHC.

 \end{abstract}

\maketitle

\newcommand{\newc}{\newcommand}
\newc{\gsim}{\lower.7ex\hbox{$\;\stackrel{\textstyle>}{\sim}\;$}}
\newc{\lsim}{\lower.7ex\hbox{$\;\stackrel{\textstyle<}{\sim}\;$}}

\def\beq{\begin{equation}}
\def\eeq{\end{equation}}
\def\beqn{\begin{eqnarray}}
\def\eeqn{\end{eqnarray}}
\def\calM{{\cal M}}
\def\calV{{\cal V}}
\def\calF{{\cal F}}
\def\half{{\textstyle{1\over 2}}}
\def\quarter{{\textstyle{1\over 4}}}
\def\ie{{\it i.e.}\/}
\def\eg{{\it e.g.}\/}
\def\etc{{\it etc}.\/}
\def\met{\displaystyle{\not}E_T}


\def\inbar{\,\vrule height1.5ex width.4pt depth0pt}
\def\IR{\relax{\rm I\kern-.18em R}}
 \font\cmss=cmss10 \font\cmsss=cmss10 at 7pt
\def\IQ{\relax{\rm I\kern-.18em Q}}
\def\IZ{\relax\ifmmode\mathchoice
 {\hbox{\cmss Z\kern-.4em Z}}{\hbox{\cmss Z\kern-.4em Z}}
 {\lower.9pt\hbox{\cmsss Z\kern-.4em Z}}
 {\lower1.2pt\hbox{\cmsss Z\kern-.4em Z}}\else{\cmss Z\kern-.4em Z}\fi}




\section{Introduction}
\label{sec:Introduction}

Exploring mechanism for Electroweak symmetry breaking (EWSB) and searching
for the Higgs boson has been the primary goal of the current and
future collider experiments.  The null results of the Large Electron-Positon Collider (LEP) Higgs
searches excluded a Standard Model (SM) Higgs boson with mass smaller  than 114.4 GeV
at 95\% C.L. \cite{lephiggsSM}.   In the SM, a global fit to the
precision measurements gives a best fit value of  $m_{H_{\rm SM}}=87^{+35}_{-26} $
GeV, with $m_{H_{\rm SM}} < 186$ GeV at 95\% C.L. \cite{lepewwg}.  There are also upper and
lower limits on the SM Higgs mass coming from the requirement of
perturbativity and vacuum stability.  At electroweak scale, those
constraints are typically weak: $50\ {\rm GeV}  \lesssim  m_{H_{\rm SM}} \lesssim 800\ {\rm GeV}$ \cite{SMhiggslimit_theory}.
Therefore, it is relatively easy to accommodate the LEP Higgs search
limit in the framework of the SM.

There are, however, indications of new physics beyond the SM.  The stabilization of the
Higgs mass at the electroweak scale requires the introduction of new physics at TeV scale.  The existence of dark matter also provides an
unambiguous evidence of new physics beyond the Standard Model.
Supersymmetry(SUSY) has been a very promising new physics candidate and
has been studied extensively in the literature \cite{susy}.  The LEP 
limit of the SM Higgs mass bound can be directly applied to the
light CP-even Higgs in the Minimal Supersymmetric Standard Model
(MSSM) in the decoupling region.  The mass of the light CP-even
Higgs in the MSSM, however, is bounded to be less than $m_Z$ at
the tree level.  It receives large radiative correction, dominantly from
top and stop loops due to its large Yukawa couplings.  The stop masses
are typically required to be large in order to push the Higgs mass
above the LEP limits, which could lead to significant
fine-tuning in electroweak symmetry breaking.  To resolve the tension
between the LEP Higgs search limits and the fine-tuning in EWSB, studies have been done in the direction of extending the minimal model  or questioning the
definition for naturalness \cite{naturalness}.
 
We could, however, also take this tension as an indication that the
data is pointing us to a particular region of the MSSM parameter
space that the LEP null results can be accommodated while the
fine-tuning in the Higgs potential could be minimized.  This
so-called ``golden" region was explored in
Ref.~\cite{perelstein}.  It is shown that such golden
region is characterized by a   small value of the $\mu$
parameter, as well as relatively small soft masses for $m_{Q_3}$ and $m_{u_3}$,
and a relatively large stop trilinear $A$-term $A_t$.  The mass
spectrum resulting from this contains light neutralinos and
charginos with a large Higgsino content, due to the small $\mu$
value.  The stop mass eigenstates $\tilde{t}_1$ and $\tilde{t}_2$ in
golden region are relatively light (sub-TeV) with a large mass
splitting.  One direct consequence of this large mass splitting is
that the decay of $\tilde{t}_2 \rightarrow \tilde{t}_1 Z$ is
kinematically allowed.  Reconstruction of the dilepton invariant
mass from $Z$ decay offers a good discrimination of the signal over
the SM backgrounds.  The collider analyses of inclusive signature
$pp\rightarrow \tilde{t}_2 \tilde{t}_2^* \rightarrow Z + 2 j_b + \met
+ X$ at the Large Hadron Collider (LHC) have been studied in Ref.~\cite{perelstein}. A 3
$\sigma$ observation could be reached with 75 ${\rm fb}^{-1}$
luminosity, while 5 $\sigma$ reach is possible with 210 ${\rm
fb}^{-1}$.

The mass parameter for $\tilde{b}_L$ in the MSSM is closely related to that of
the $\tilde{t}_L$, both determined dominantly by $m_{Q_3}$.
Therefore, in the SUSY golden region, a light sbottom is also within
the reach of the LHC.  In our analyses, we studied the collider
signature of the light sbottom pair production in the SUSY golden region.  The light sbottom decays mostly into $t \chi_1^-$ or 
$W^- \tilde{t}_1 $.  There are typically two $b$ jets
plus two $W$'s in the sbottom pair production decay final states. In this paper, we
studied events with $2 j_b+  2\ {\rm jets}+   1\ {\rm lepton}+{\met} +X$ at the  LHC  and
developed a set of cuts to identify the signal and suppress the
backgrounds.

Our analyses differ from the conventional SUSY searches for the sbottom, in which the light 
$\tilde{b}_1$ (mostly a $\tilde{b}_L$) decays via 
$\tilde{b}_1 \rightarrow b \chi_2^0$ with $\chi_2^0 \rightarrow \chi_1^0 \ell^+ \ell^-$ \cite{ATLASTDR, CMSTDR}, for $\chi_2^0$ being mostly a wino.  
In the SUSY golden region, the branching ratio of $\tilde{b}_1 \rightarrow b \chi_2^0$ is small since 
$\chi_2^0$ is mostly Higgsino and $\tilde{b}_1 -b -\chi_2^0$ coupling is suppressed by the small bottom Yukawa coupling.  
Channel of $\tilde{b}_1 \rightarrow t \chi_1^-$ with hadronic top decay has also been studied \cite{ATLASTDR}.
Reconstruction of the hadronic top events provide a good discrimination over the SM backgrounds. 
In our study, $\tilde{b}_1$ decays via either $t \chi_1^-$ or $W^-\tilde{t}_1 $ with $\chi_1^\pm$ and $\tilde{t}_1$ further decays.  $bW$  in the final state does not necessary come from an on-shell top.  In addition,  we require one $W$ decay leptonically and one $W$ decay hadronically and study its LHC discovery reach of such semileptonic channel.

In Sec.~\ref{sec:goldenregion}, we discuss
the SUSY golden region and present the benchmark point
that is used in the  current analyses.  In
Sec.~\ref{sec:collider}, we present the details of the collider
analyses and the discovery reach at the LHC. In Sec.~\ref{sec:conclusion}, we
conclude.

\section{SUSY Golden Region}
\label{sec:goldenregion}

Naturalness and experimental Higgs mass bound point to a ``golden region" in the MSSM parameter space.   Since the Higgs sector couples strongly to the top/stop sector, while weakly to the rest of the MSSM, we focus on  the Higgs and top/stop sectors, which are determined by seven  parameters: $m_{H_u}^2, \ m_{H_d}^2,\ \mu, \ b, \ m_{Q_3}^2, m_{u_3}^2$ and $A_t$.  After electroweak symmetry breaking,  the neutral components of the Higgs fields obtain vacuum expectation values as $\langle H_u^0 \rangle = v_u$ and $\langle H_d^0 \rangle = v_d$ with $\sqrt{v_u^2+v_d^2}=174$ GeV.  Define $\tan\beta=v_u/v_d$ and replace one of the Higgs mass parameters by the CP-odd Higgs mass $m_A$, we are left with  six independent parameters:
\begin{equation}
\tan\beta, \mu, m_A, m_{Q_3}^2, m_{u_3}^2,  A_t.
\end{equation}
Naturalness consideration in the tree-level electroweak symmetry breaking relations leads to \cite{perelstein} 
\begin{equation}
\frac{\mu}{m_Z} < \frac{\Delta^{1/2}}{2},\ \ \ 
\frac{m_A}{m_Z}<\frac{\Delta^{1/2}}{2}\tan\beta,
\end{equation}
where $\Delta \leq 100$ corresponds to  a fine tuning of 1\% or better. 
Including the quantum corrections to the Higgs potential from the stop sector further constrains the stop mass eigenvalues and mixing angle $({m}_{\tilde{t}_1}, {m}_{\tilde{t}_2}, \theta_{\tilde{t}})$, in which relatively  light masses   are preferred to minimize the fine-tuning.

The Higgs searches at the LEP gave a lower limit on the SM Higgs mass as \cite{lephiggsSM}
\begin{equation}
m_{H_{\rm SM}}\gtrsim 114.4\  {\rm GeV},
\end{equation}
which can also be applied to the light CP-even Higgs boson in the decoupling region of the MSSM parameter space.  Although MSSM could accommodate a lighter Higgs (around 90 GeV) \cite{lephiggsMSSM} while still being consistent with the LEP Higgs search results, it only happens in a restricted region of the MSSM parameter space, which can be viewed as additional source of fine-tuning.  In Ref. \cite{perelstein}, authors used the LEP limit of 114.4 GeV as a lower bound on the light MSSM CP-even Higgs mass.  The dominate loop contribution to the mass of the light CP-even Higgs boson comes from the stop sector.  Accommodating the LEP Higgs search bound requires heavier stop masses and/or large left-right stop mixing.  Taking into account the collider search limit on sparticle masses, constraints from the $\rho$ parameter, rare decays $b\rightarrow s \gamma$, as well as minimizing the fine tuning, we are  limited to the MSSM golden region with small $\mu$ and $m_A$, relatively small value for $m_{Q_3}^2, m_{u_3}^2$ and large value for $A_t$.  

In our analyses, we used the benchmark point chosen in Ref.~\cite{perelstein}.  
The MSSM input parameters at the weak scale are given in Table~\ref{tab:benchmark}.
We assume there is no flavor off-diagonal terms and all the tri-linear $A$ terms are zero 
except $A_t$.   
 All the gaugino masses $M_{1,2,3}$ as well the masses for the slepton and first two generation squarks  are chosen to be 1 TeV.  The mass parameter for the $\tilde{b}_R$, $m_{d_3}$, is also chosen to be heavy.  Varying those parameters does not have significant effects on the Higgs potential, as well as the light sbottom (mostly $\tilde{b}_L$ in our case) pair production channel that we consider below.  

\begin{table}
\begin{tabular}{|c|c|c|c|c|c|c|c|c|c|c|c|} \hline
$m_{Q_3}$&$m_{u_3}$ &$m_{d_3}$&$A_t$&$\mu$&$m_A$&$\tan\beta$&
$M_1$&$M_2$&$M_3$&$m_{\tilde{q}}$&$m_{\tilde{l}}$
\\\hline 548.7&547.3&1000&1019&250&200&10&1000&1000&1000&1000&1000\\
\hline
\end{tabular} \\
\caption{MSSM input parameters defined at the
weak scale for the benchmark point of the MSSM golden region, taken from Ref.\cite{perelstein}. All dimensionful parameters are in unit of GeV.}
\label{tab:benchmark}
\end{table}

\begin{table}
\begin{tabular}{|c|c|c|c|c|c|c|c|c|c|} \hline
$m_{\tilde{t}_1}$&$m_{\tilde{t}_2}
$&$m_{\tilde{b}_1}$&$m_{{\chi}_1^0}$&$m_{{\chi}_2^0}$&$m_{\chi_1^{\pm}}$&$m_{h^0}$&$m_{H^0}$&$m_{A}$&$m_{H^{\pm}}$
\\ \hline 
398&688&550&243&253&247&118&201&200&216\\ \hline
\end{tabular} \\
\caption{Physical spectrum of  light sparticles for the benchmark point in the MSSM golden region, in unit of GeV.} \label{tab:mass}
\end{table}

The physical mass spectrum of light particles for the benchmark point is given in Table~\ref{tab:mass}, which is obtained using  SOFTSUSY 2.0.11\cite{softsusy}.  
With the small value of $\mu$, the two lightest neutralinos and charginos are almost degenerate, which are mostly Higgsinos.  
The mass for the light CP-even Higgs is about 118 GeV, above the LEP Higgs search limit.
The heavy CP-even Higgs, as well as the CP-odd Higgs and the charged Higgses are around 200 GeV.   
Due to the large left-right mixing in the stop sector, there is a large mass splitting in the two stop mass eigenstates: $m_{\tilde{t}_2}-m_{\tilde{t}_1}>m_Z$, which is a generic feature of  the MSSM golden region.   Utilizing this feature,
Ref.~\cite{perelstein} studied the process of $\tilde{t}_2\tilde{t}_2^*$ pair production at the LHC with at least one $\tilde{t}_2$ decays via $\tilde{t}_2 \rightarrow \tilde{t}_1 Z$.  Inclusive signature of  $Z$ + 2 $b$-jets + $\met$ + $X$ is analyzed, where $Z$ decays leptonically  into pair of electrons or muons.  It is found that by requiring the reconstruction of the lepton pair around the $Z$ peak, a large $p_T$ cut on the first two leading jets with at least one $b$-tagging, a large boost factor of $Z$ boson and a large $\met$ cut, a 3 $\sigma$ observation of this inclusive signal is possible with 75 ${\rm fb}^{-1}$ luminosity, while a 5 $\sigma$ discovery is possible with 210 ${\rm fb}^{-1}$ data. 

Ignoring the left-right mixing in the sbottom sector, the mass of $\tilde{b}_L$ is also determined by 
$m_{Q_3}$.  Therefore, in the MSSM golden region, one of the sbottom is also relatively light, which can be copiously produced at the LHC.  For the benchmark point presented in Table~\ref{tab:benchmark}, the mass of the light sbottom is 550 GeV.  The leading order pair production cross section at the 14 TeV LHC is about 214 fb.  The light sbottom dominantly decays into $t \chi_1^- $ and $W^-\tilde{t}_1 $, with branching ratio of 51.4\% and 46.5\%, respectively.  $\chi_1^\pm$ decays into $\chi_1^0$ with soft jets and leptons, due to the small mass splittings between charged and neutral Higgsinos.  $\tilde{t}_1$ dominantly decays into $b \chi_1^+$, with $ \chi_1^+$ further decays.    Therefore,  the decay products of $\tilde{b}_1$ include at least one $b$-jet plus $W$ plus $\chi_1^0$.    In our analyses below, we consider the pair production of $\tilde{b}_1 \tilde{b}_1^*$ at the LHC with $\sqrt{s}=14$ TeV.  
Demanding one $W$ decay leptonically and one $W$ decay hadronically, we study the collider signature of 
\begin{equation}
pp \to \tilde b_1 {\tilde b_1}^* \to 2 j_b+ \geq 2\ {\rm jets}+ \geq 1\ {\rm lepton}+{\met}.
\label{eq:decay}
\end{equation} 

Note that for the parameter choices of the benchmark point in Table~\ref{tab:benchmark}, $M_{1,2,3}$ are taken to be very heavy and the light neutralinos $\chi_{1,2}^0$ and charginos $\chi_1^\pm$ are mostly Higgsinos.  For smaller value of $M_1$, the decay branching ratios of $\tilde{b}_1$ do not change much since $\tilde{b}_1 \rightarrow b \chi_i^0$ is suppressed by either the small U(1) gauge coupling or the small bottom Yukawa coupling.   For smaller value of $M_2$, channels of $\tilde{b}_1 \rightarrow t \chi_{1,2}^\pm$ both open up, which do not change the collider signature of Eq.~(\ref{eq:decay}) given the further decay of $\chi_{1,2}^\pm$.  
$\tilde{b}_1$ could also decay into $b \chi_i^0$ with sizable branching ratio, where $\chi_i^0$ is mostly wino.
Our results below could be rescaled by the branching ratio for such case of small $M_2$.

\section{Collider Analyses}
\label{sec:collider}

We generate the event samples   for the signal process  
 at parton-level using the MadGraph 4.4.26~\cite{MadGraph}
package.  These events were subsequently passed through PYTHIA 6.420~\cite{PYTHIA} 
for parton showering and hadronization, and then through PGS4~\cite{pgs} to 
simulate the effects of a realistic detector.  The corresponding total leading order cross section 
for sbottom pair production at the 14 TeV LHC 
is estimated to be about
$214$ fb. 
The dominating SM background comes from $t\bar{t}$, with one $W$ decay leptonically and the 
other decay hadronically.    Another irreducible background is $t\bar{t}Z$, with $Z\rightarrow \nu\bar{\nu}$, mimicking the missing energy signature from the lightest neutralino.  
Other possible backgrounds are 
$t\bar{t}W$, $WZjj$, $WWjj$ and $Wjjjj$, with $j$ being light quarks.  All the background events are generated similar to the signal process, except $Wjjjj$, which is generated using ALPGEN 2.13\cite{alpgen}.

The first set of cuts (referred to Cut I) is designed to mimic a realistic
detector acceptance:
\begin{itemize}
\item  At least four jets with  $|\eta_j|<3$ and $p_{Tj1}>20$ GeV for the leading jets and 
 $p_{Tj}>15$ GeV for the other three jets.
\item At least one charged lepton (electron or muon) with $p_T^\ell>15$~GeV and $|\eta_\ell|<2.4$.
\item For jet and lepton isolation, we require $\Delta R_{jj} > 0.4$ for each possible 
  jet pairing, and $\Delta R_{j\ell}>0.4$ for each combination of one 
   jet and one charged lepton.
\end{itemize}    

\begin{figure}[bht]
\begin{center}
\resizebox{3.15in}{!}{\includegraphics*{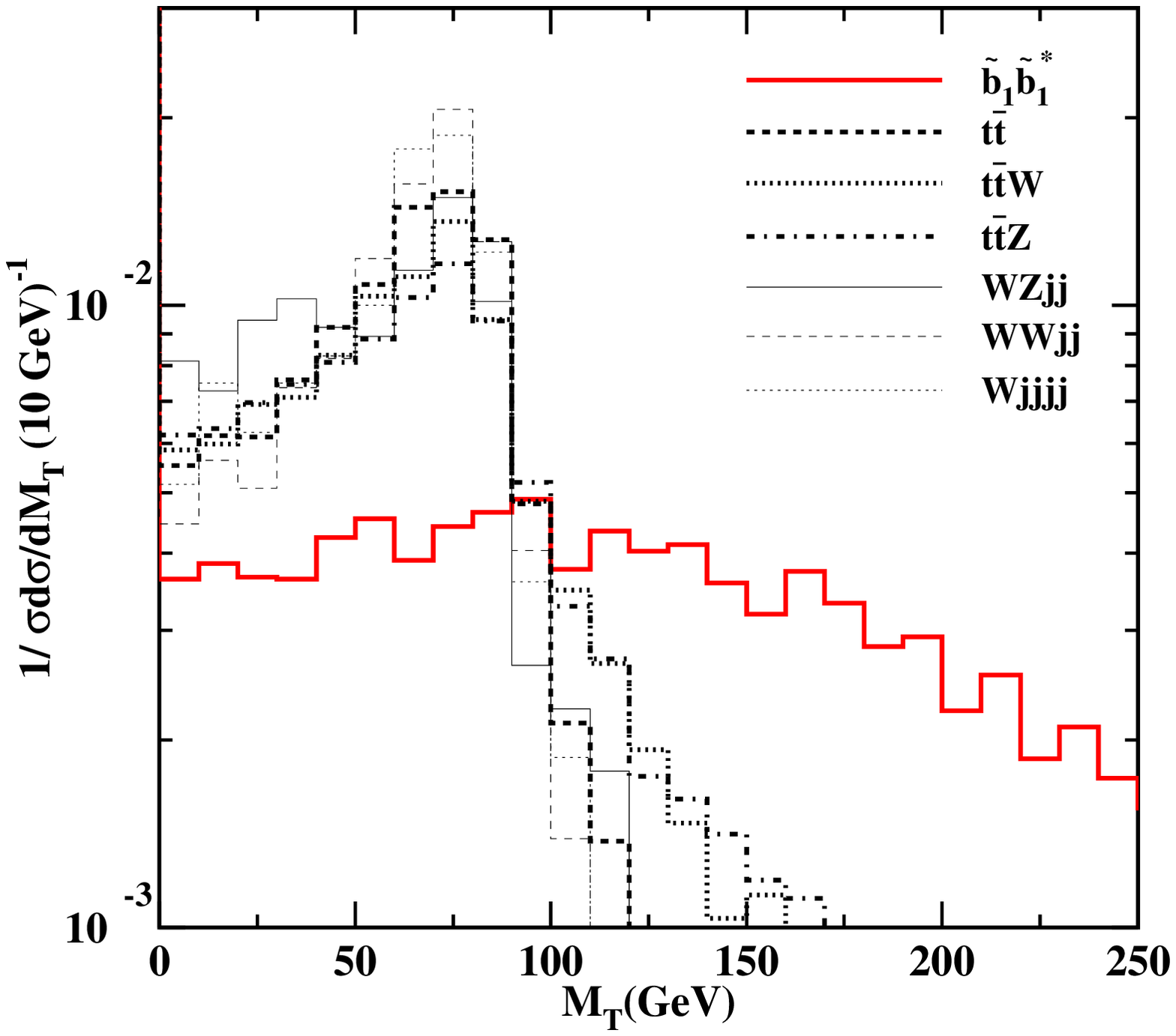}}
\resizebox{3.15in}{!}{\includegraphics*{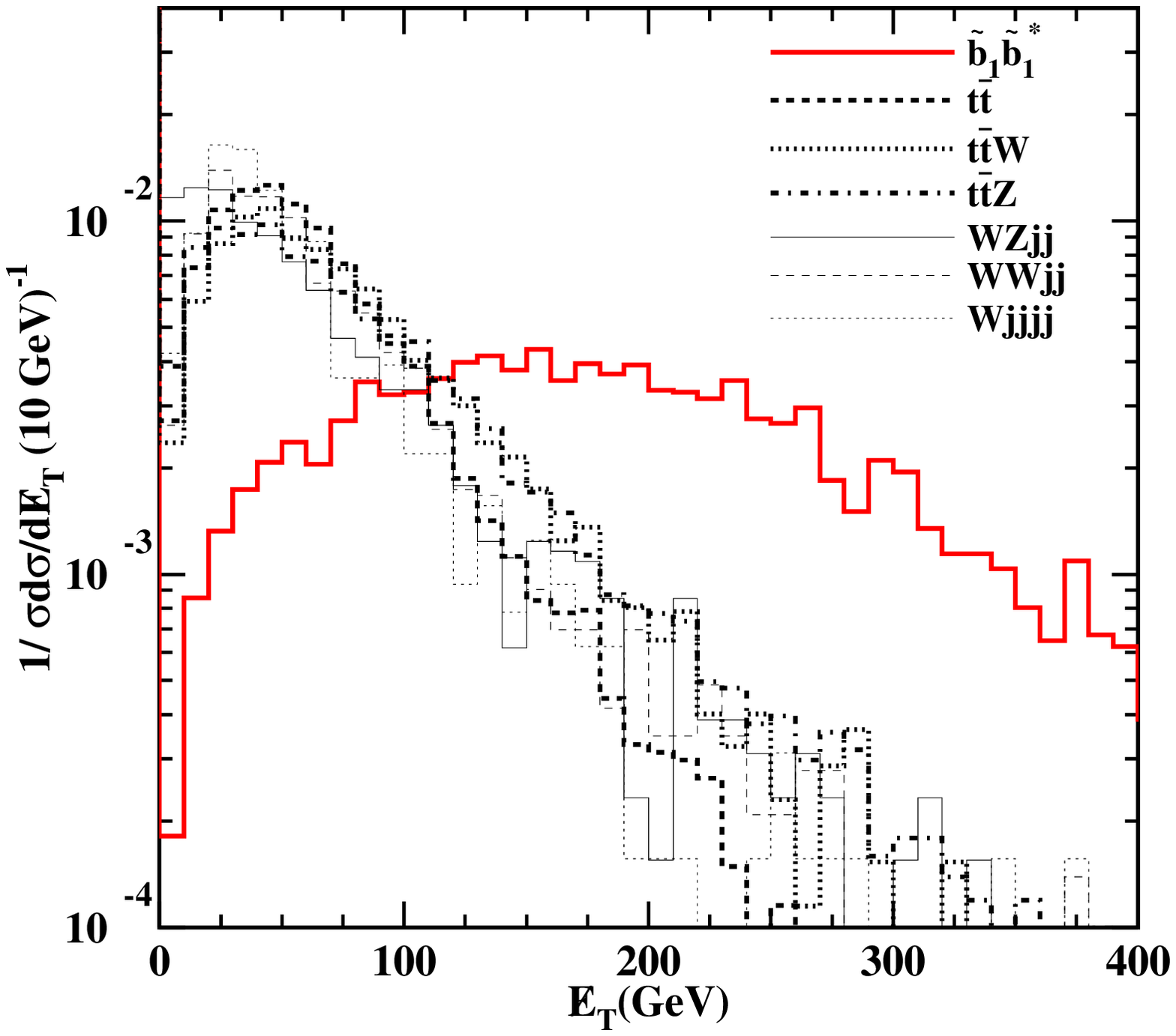}}
\caption{Transverse mass $M_T$ (left panel) and missing transverse energy $\met$ (right panel) distributions for the signal and SM backgrounds.  
}
\label{fig:met_mT}
\end{center}
\end{figure}

 For    SM backgrounds $t\bar{t}$, $WWjj$, $WZjj$ and $Wjjjj$,  
both the  lepton and missing $E_T$ come from leptonic $W$ decay, 
We define a transverse-mass variable $M_{T}$, 
\begin{equation}
  M_{T}^2 \equiv (E_{\ell} + \displaystyle{\not}E_T)^2 - 
             (\vec{p}_{T\ell} + \displaystyle{\not}\vec{p}_T)^2,
\end{equation}
where $\displaystyle{\not}E_T$ and $\displaystyle{\not}\vec{p}_T$  denote the
total missing transverse energy and missing transverse momentum vector, respectively. 
The distribution for $M_{T}$ drops sharply around $m_W$ for the SM $W(\ell\nu)$ backgrounds, as shown in the left panel of Fig.\ref{fig:met_mT}.  
For the signal process, on the other hand, due to the additional contribution to the missing $E_T$ from $\chi_1^0$, $M_T$ drops more gently beyond $m_W$.  For the background SM processes $ttZ$ and $ttW$ with additional neutrinos from $W$ or $Z$ decay, $M_T$   extends beyond $m_W$ as well.  Adopting a cut of 
\begin{itemize}
\item{Cut II: $M_T > 120$ GeV.}
\end{itemize}
greatly reduces the  $t\bar{t}$, $WWjj$, $WZjj$ and $Wjjjj$ backgrounds.

The right panel of Fig.\ref{fig:met_mT} shows the $\met$ distribution of signal and backgrounds.  It is clear that the signal process has a larger missing $E_T$ due to the presence of two $\chi_1^0$, while the missing $E_T$ in the SM processes are typically small.  Therefore, we adopt a cut of 
\begin{itemize}
\item{ Cut III: $\met> 225$ GeV.  }
\end{itemize}
to further suppress the SM backgrounds.

\begin{table}
\begin{center}
\begin{tabular}{|c|c|c|c|c|c|c||c|c|} \hline
process &
$\sigma_i$(fb)&$N_{total}$&cut-I(\%)&cut-II(\%)&cut-III(\%)&$\sigma_f^{III}$(fb)&cut-IV(\%)&$\sigma_f^{IV}$(fb)
\\\hline 
$\tilde{b}_1{\tilde{b}_1}^{*}$&$2.14 \times 10^2$&  40000&  19&   51& 44&9.1&35& 3.2  \\ \hline 
$ t\bar t$ & $5.38 \times 10^5$&   1826954& 15& 3.8& 2.6&77&33&26\\ \hline 
$ t\bar t W$ &$5.22 \times 10^2$&120161& 26& 11& 8.8&1.3&37& 0.5\\ \hline 
$ t\bar t Z$ &$6.85 \times 10^2$&166420& 27&   13&    12&2.9& 37&1.1\\ \hline 
$WZjj$& $3.90 \times 10^4$& 1259561  &5.6 &4.0   &14    & 12&6.4 &0.8\\ \hline 
$ WWjj$&$6.88 \times 10^4$& 1260060  &6.3 & 2.8  &8.3    &10 &4.9&0.5\\ \hline 
$Wjjjj$&$2.55 \times 10^7$ &2512703  & 2.8& 0.9& 5.0&315& 3.2&10 \\\hline
\multicolumn{5}{c|} {}&
$S/\sqrt{B}$&4.5&$S/\sqrt{B}$&5.1 \\ \cline{6-9}
\multicolumn{5}{c|} {}&
$S/B$&0.02&$S/B$&0.08 \\ \cline{6-9}
\end{tabular} 
\caption{Summary of the cross sections and cut efficiencies (with respect to the previous level of cut) for the signal and background processes at the 14 TeV LHC before and after each cut. The $WZjj$, $WWjj$ and $Wjjjj$ cross section before the cuts are calculated with a precut of $p_{Tj}>10$ GeV, $|\eta_j|<5$ and $\Delta R_{jj} > 0.2$.  The third column shows the total number of events that is simulated for each process.}
\label{tab:crosssection}
\end{center}
\end{table}

Table~\ref{tab:crosssection} summarize the signal and background cross sections, as well as the  cut efficiencies.   After three levels of cuts, the remaining dominant background is $Wjjjj$, followed by $t\bar{t}$.
For  an integrated luminosity of $100\ {\rm fb^{-1}}$, about
900  $\tilde{b}_1{\tilde{b}_1}^*$ events can be found at
LHC after Cuts-I, II and III.  The significance is about $S/\sqrt{B}=4.5$ for ${\cal L}=100\ {\rm fb}^{-1}$ with $S/B=0.02$.  

To further reduce the $Wjjjj$ background with non-$b$-jets, we can demand $b$-tagging on at least
one jet: 
\begin{itemize}
\item Cut IV,
at least one $b$-tagging in the two leading  jets.
\end{itemize}
The resulting cross sections after this cut are shown in the last two columns of 
Table~\ref{tab:crosssection}.  
Backgrounds $Wjjjj$, $WZjj$ and $WWjj$ are reduced greatly with $b$-tagging.
$t\bar{t}$ background, however, now becomes dominant, which   leads to a final 
significance level\footnote{For $Wjjjj$ process, there is only one event (out of 2,512,703
total number of events)
left after all the cuts.
Using Poisson statistics, we would expect an upper limit of 4.74 events at 95$\%$ C.L., which leads to a $Wjjjj$ cross section of 48 fb.  Using this 95$\%$ C.L. upper limit, the resulting final significance level $S/\sqrt{B}$ is 3.6 for 100 ${\rm fb}^{-1}$ integrated luminosity with $S/B=0.04$.}
of $S/\sqrt{B}=5.1$ with $S/B=0.08$ for 100 ${\rm fb}^{-1}$ integrated luminosity.     Imposing such $b$-tagging could be used to suppress other SUSY processes with non-$b$-jets that have similar signatures.

All the cross sections that are presented above are leading-order results.  The $K$-factor for the dominating $t\bar{t}$ background is about 1.41 $-$ 1.65 \cite{NLOtt}.  For the signal process, the $K$-factor depends on the sbottom masses.  For a 400 GeV sbottom, it is about 1.40 \cite{NLOsbsb}.  The discovery significance level will not change much if the next leading order QCD corrections for both signal and background processes are included.

The signal that we discussed above, $pp\rightarrow \tilde{b}_1 \tilde{b}_1^* \rightarrow  2 j_b +  2\  {\rm jets} +   1\ {\rm lepton} + \met +X$, is complementary to the process of 
$pp \rightarrow \tilde{t}_2 \tilde{t}_2^* \rightarrow Z + 2 j_b + \met + X$ that was discussed in 
Ref.~\cite{perelstein}.    Observation of both processes at the LHC could be a strong indication of 
the MSSM golden region.  Another related process is the pair production of $\tilde{t}_2 \tilde{t}_2^*$, with $\tilde{t}_2 \rightarrow \tilde{b}_1 W$.  With the additional $bW$ from $\tilde{b}_1$ decay, such pair production process could have four $W$s in the final states.  This could be another complementary process for the MSSM golden region.   Another interesting final state could be one $\tilde{t}_2$ decay via $Z\tilde{t}_1$ while the other decay via $W\tilde{b}_1$.  Such process  could also be useful in identifying the MSSM golden region.

\section{Conclusion}
\label{sec:conclusion}

Naturalness and Higgs search limit at the LEP points us to the SUSY golden region where the Higgs mass bound is satisfied while the fine tuning in the electroweak symmetry breaking is minimized.  The characteristics of this region include small value for $\mu$ and $m_A$, moderate value for the stop masses and a large left-right mixing in the stop sector.  Ref.~\cite{perelstein} studied the 
pair production of $\tilde{t}_2\tilde{t}_2^*$ with at least one $\tilde{t}_2$ decay via 
$\tilde{t}_2 \rightarrow \tilde{t}_1 Z$, motivated by the large mass splitting of 
$m_{\tilde{t}_2} - m_{\tilde{t}_1}$.   Since the mass of the $\tilde{b}_L$ is also determined by 
$m_{Q_3}$, the light sbottom can also be copiously produced at the LHC.  In this paper, we studied the sbottom pair production and its consequent decay via $t \chi_1^\pm$ and $W \tilde{t}_1$.  The collider signature is 2 $b$-jets +   2 jets + 1 lepton + $\met$ + $X$.  
We found that   a 5 $\sigma$ discovery could be reached with 100 ${\rm fb^{-1}}$ integrated luminosity.
Observation of both signatures at the LHC could be a strong indication for the existence of the SUSY golden region.  With more luminosities, we could further pin down the mass differences of the light particle spectrum by studying various distributions of final decay products.

Our above analyses for the sbottom pair production and decay is performed for one particular benchmark point in the MSSM golden region as defined in Table~\ref{tab:benchmark}.  
It is quite robust if the light spectrum only includes Higgsinos, stops and sbottoms.  Lowering the other particle masses, for example,  $M_2$, could change the decay pattern of the light sbottom and lead to different signature.  In that case, conventional sbottom search strategy \cite{ATLASTDR, CMSTDR} could be used in addition to the signature  that is discussed in this paper.

\section{Acknowledgments}

We would like to thank M. Perelstein  for useful discussions. 
We would also like to thank INPAC at Shanghai Jiaotong University 
for its hospitality while part of the work is finished. 
HL and ZG are supported in part by NSFC and Natural Science Foundation of Shandong Province (JQ200902). SS is supported by the Department of Energy
under Grant~DE-FG02-04ER-41298.


\smallskip



\begin{references}

\bibitem{lephiggsSM}
  R.~Barate {\it et al.}  [LEP Working Group for Higgs boson searches and
                  ALEPH Collaboration and  DELPHI Collaboration and L3 Collaboration and OPAL Collaboration],
  Phys.\ Lett.\  B {\bf 565}, 61 (2003)
  [arXiv:hep-ex/0306033].
  
  
\bibitem{lepewwg}
The LEP Electroweak Working Group,   0911.2604.
  Write-up LEPEWWG/2009-01 CERN-PH-EP/2008--23 arXiv:0911.2604 [hep-ex] (November 2009).
  
  \bibitem{SMhiggslimit_theory}
G.~Altarelli and G.~Isidori,
  Phys.\ Lett.\  B {\bf 337}, 141 (1994);
J.~A.~Casas, J.~R.~Espinosa and M.~Quiros,
  Phys.\ Lett.\  B {\bf 342}, 171 (1995)
  [arXiv:hep-ph/9409458];
  T.~Hambye and K.~Riesselmann,
  arXiv:hep-ph/9708416;
 G.~Isidori, G.~Ridolfi and A.~Strumia,
  Nucl.\ Phys.\  B {\bf 609}, 387 (2001)
  [arXiv:hep-ph/0104016].
  
 
 
 
 \bibitem{susy}  
 S.~P.~Martin,
  arXiv:hep-ph/9709356.

\bibitem{naturalness} 
G.~F.~Giudice and R.~Rattazzi,
  Nucl.\ Phys.\  B {\bf 757}, 19 (2006)
  [arXiv:hep-ph/0606105].
  

\bibitem{perelstein}
 M.~Perelstein and C.~Spethmann,
  JHEP {\bf 0704}, 070 (2007)
  [arXiv:hep-ph/0702038].
  


\bibitem{ATLASTDR}
 ATLAS detector and physics performance, Technical design report,  Vol. 2.

\bibitem{CMSTDR}  G.~L.~Bayatian {\it et al.}  [CMS Collaboration],
  J.\ Phys.\ G {\bf 34}, 995 (2007).

  \bibitem{lephiggsMSSM}
  S.~Schael {\it et al.}  [ALEPH Collaboration and DELPHI Collaboration and
                  L3 Collaboration and ],
  Eur.\ Phys.\ J.\  C {\bf 47}, 547 (2006)
  [arXiv:hep-ex/0602042].



\bibitem{softsusy}B.~C.~Allanach,
  Comput.\ Phys.\ Commun.\  {\bf 143}, 305 (2002)
  [arXiv:hep-ph/0104145].


\bibitem{MadGraph}
  J.~Alwall {\it et al.},
  JHEP {\bf 0709}, 028 (2007)
  [arXiv:0706.2334 [hep-ph]].

\bibitem{PYTHIA}
  T.~Sjostrand, S.~Mrenna and P.~Skands,
  JHEP {\bf 0605}, 026 (2006)
  [arXiv:hep-ph/0603175].

\bibitem{pgs}
``PGS -- Pretty Good Simulator'',\\ {\tt
  http://www.physics.ucdavis.edu/}
  $\sim${\tt conway/research/software/pgs/pgs4-general.html}



\bibitem{alpgen}  
F.~Caravaglios, M.~L.~Mangano, M.~Moretti and R.~Pittau,
  Nucl.\ Phys.\  B {\bf 539}, 215 (1999)
  [arXiv:hep-ph/9807570];
M.~L.~Mangano, M.~Moretti and R.~Pittau,
  Nucl.\ Phys.\  B {\bf 632}, 343 (2002)
  [arXiv:hep-ph/0108069];
  M.~L.~Mangano, M.~Moretti, F.~Piccinini, R.~Pittau and A.~D.~Polosa,
  JHEP {\bf 0307}, 001 (2003)
  [arXiv:hep-ph/0206293].
  
\bibitem{NLOtt}M.~Cacciari, S.~Frixione, M.~L.~Mangano, P.~Nason and G.~Ridolfi,
  JHEP {\bf 0809}, 127 (2008)
  [arXiv:0804.2800 [hep-ph]];
  W.~Bernreuther and Z.~G.~Si,
  Nucl.\ Phys.\  B {\bf 837}, 90 (2010)
  [arXiv:1003.3926 [hep-ph]].

\bibitem{NLOsbsb} 
  W.~Beenakker, S.~Brensing, M.~Kramer, A.~Kulesza, E.~Laenen and I.~Niessen,
  JHEP {\bf 1008}, 098 (2010)
  [arXiv:1006.4771 [hep-ph]].


  
\end{references}
\end{document}